\documentclass[10pt,conference,a4paper]{IEEEtran}

\usepackage[utf8]{inputenc}

\usepackage{blindtext, graphicx}
\usepackage[utf8]{inputenc}
\usepackage[english]{babel}
\usepackage{csquotes}
\usepackage{wrapfig}
\usepackage{float}
\usepackage{verbatim}
\usepackage{subcaption}
\usepackage[font=footnotesize,labelfont=bf]{caption}
\usepackage{import}
\usepackage{dirtytalk}
\usepackage{subfiles}
\usepackage{array}
\usepackage{multirow}

\usepackage{lscape}
\usepackage{algorithm}
\usepackage{algorithmic}
\usepackage{rotating}
\usepackage[table,xcdraw]{xcolor}
\usepackage{longtable} % Para tablas grandes
\usepackage{array} % Para definir anchos de columnas
\usepackage{caption} % Para mejorar el formato de leyendas
\usepackage{graphicx} % Permite escalado si es necesario
\usepackage{xcolor, colortbl} % Para resaltar si es necesario
\usepackage{booktabs} % Para mejorar el diseño de la tabla
\usepackage{tabularx} % Para que la tabla use todo el ancho disponible
\usepackage{amsmath}
\usepackage{times}
\DeclareGraphicsExtensions{.png,.eps,.ps,.pdf}
\usepackage{url}
\usepackage{hyperref}
\usepackage{xurl}
\usepackage[most]{tcolorbox}
\usepackage{multirow}
\usepackage{tabularx}
\usepackage{makecell}
\usepackage{svg}
\usepackage{pifont}
\definecolor{ao}{rgb}{0.0, 0.5, 0.0}
\definecolor{amber}{rgb}{1.0, 0.49, 0.0}

\let\labelindent\relax
\usepackage{enumitem}

\usepackage{lscape} 

\graphicspath{{Figures/}}

\usepackage{tikz} % For \foreach used for Orcid icon
\newcommand{\orcidicon}{\includegraphics[width=0.32cm]{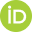}}
\foreach \x in {A, ..., Z}{%
\expandafter\xdef\csname orcid\x\endcsname{\noexpand\href{https://orcid.org/\csname orcidauthor\x\endcsname}{\noexpand\orcidicon}}
}

\begin{document}

\title{MARCIM-WG: A cyber wargame proposal based on math modeling applied in a naval scenario}

\author{\IEEEauthorblockN{
\orcidA{}Diego Cabuya-Padilla$^{1}$,
\orcidB{}Daniel D\'iaz-L\'opez$^{2,3}$,
\orcidC{}Carlos Castaneda-Marroqu\'in$^{4}$
}

\IEEEauthorblockA{$^1$  `Poseidón'' research group, Naval Academy of Strategic Studies, Bogot\'a, Colombia \\ diego.cabuya@enap.edu.co}
\IEEEauthorblockA{$^2$Department of Information and Communications Engineering, University of Murcia, 30100, Murcia, Spain\\
danielorlando.diaz@um.es}
\IEEEauthorblockA{$^3$School of Engineering, Science and Technology, Universidad del Rosario, Bogot\'a, Colombia\\
danielo.diaz@urosario.edu.co}
\IEEEauthorblockA{$^4$`GICCDN'' research group, "Almirante Padilla'' Naval Cadet Academy, Cartagena, Colombia\\ carlos.castaneda@enap.edu.co}
}

\maketitle
\begin{abstract}
As maritime operations increasingly depend on interconnected digital ecosystems, cyber incidents can propagate across maritime networks and degrade critical services. Strengthening strategic Cyber Situational Awareness (CSA) therefore requires training mechanisms that expose decision-makers to evolving attack dynamics, constrained resources, and the need to align actions with incident-response procedures. This paper introduces MARCIM-WG, a learning-oriented maritime cyberdefense wargame designed following the NATO wargaming methodology and implemented as a hybrid tabletop experience combining a physical board (tokens, indicators, and special cards) with analytically-assisted adjudication supported by a computational simulation model. The proposal is specified through High-Level Design (HLD) and Low-Level Design (LLD) specifications and instantiated in a fictional maritime cyber crisis scenario to enable structured decision cycles, friction, and measurable consequences. Validation combines (i) an operational scenario-based assessment under three configurations (pessimistic, neutral/most likely, optimistic) to verify decision sensitivity and outcome coherence, and (ii) a CSA competency and learning-outcome evaluation using a comparative design against an equivalent control group. Results show a +34.0 percentage-point improvement in the intervention group, with the largest gains in comprehension-related competencies.
\end{abstract}


{\bf Contribution type:}  {\it  Training and educational innovation}

\section{Introduction}\label{intro}

The maritime sector concentrates assets and services that sustain global trade and national security, while increasingly relying on interconnected digital ecosystems. This dependency expands the attack surface and increases the likelihood that cyber incidents translate into operational disruption, economic impact, and safety risks across ships, ports, and supporting logistics chains \cite{unctad2024review,alcaide2020critical,symes2024the,mrakovi2019maritime}. From a strategic cyberdefense viewpoint, the challenge is not limited to technical containment: leaders must preserve mission continuity under uncertainty, align stakeholders, and anticipate cascading effects across interdependent maritime services \cite{karim2022maritime,mrakovi2019maritime}.

Cyber Situational Awareness (CSA) is pivotal in that context because it supports (i) the perception of relevant cues, (ii) the comprehension of their meaning in context, and (iii) the projection of plausible future states to guide decisions in dynamic environments \cite{endsley1995toward,franke2014cyber}. In cyber crises, CSA is operationalized through the ability to translate technical signals into decision-relevant understanding and to prioritize actions under time pressure \cite{paul2013a}. In naval systems, achieving actionable awareness requires domain-grounded interpretation of evolving cyber conditions to support timely decisions \cite{jacq2019cyber}.

Wargames are a practical mechanism to strengthen decision-making in complex, adversarial environments by combining narrative, friction, and iterative learning. In cyberdefense, wargaming enables participants to rehearse decisions, explore trade-offs, and internalize procedures while preserving the essential dynamics of an adaptive adversary \cite{bodeau2018cyber,curry2018cyber,weiner1959the}. However, cyber wargames are often constrained by limited validation and uneven contextualization \cite{onduto2021implementing}. In particular, the specialized use of wargaming to foster CSA within maritime cyberdefense remains scarcely documented, despite the recognized relevance of maritime cyber risk and the growth of maritime cybersecurity research \cite{cabuyapadilla2024dyna,cabuyapadilla2024ciberseguridad,cabuyapadilla2025hybrid}.

To address this gap, this paper presents MARCIM-WG, a maritime cyberdefense wargame designed to foster strategic CSA by enabling participants to appropriate crisis-response procedures and protocols through a hybrid experience that combines a physical board with computational adjudication. The objective of MARCIM-WG is to develop CSA competencies---perception, comprehension, and projection--- \cite{endsley1995toward,franke2014cyber} by exposing decision-makers to evolving attack dynamics, resource constraints, and decision--consequence cycles within a fictional cyber crisis . The wargame is conceived and specified following the NATO wargaming methodology \cite{alliedcommandtransformation2023nato}, progressing from purpose and learning objectives to high-level and low-level specifications, scenario, execution, and results-oriented post-analysis. MARCIM-WG builds on a modeling-and-simulation foundation for maritime cyberdefense and leverages an underlying propagation model to represent crisis evolution and the effects of defensive decisions \cite{cabuyapadilla2025serdux,cabuyapadilla2025hybrid}.
Building upon these previous developments, this work extends the SERDUX-MARCIM framework by translating its mathematical and simulation capabilities into an operational wargaming environment. In this sense, MARCIM-WG represents a methodological and practical evolution, enabling the application of the model in structured decision-making scenarios aimed at developing Cyber Situational Awareness at the strategic level.

The main contributions of this work are as follows:

\begin{itemize}
\item A domain-grounded maritime cyberdefense wargame (MARCIM-WG) that explicitly operationalizes Cyber Situational Awareness (CSA) through structured decision cycles, incorporating perception, comprehension, and projection as measurable competencies within a controlled experimental environment.

\item The integration of a hybrid tabletop wargaming environment with an analytically-assisted adjudication mechanism, where player decisions are systematically translated into quantitative system-level effects through a computational simulation model, ensuring traceability between actions and outcomes.

\item A complete and methodologically consistent specification of the wargame through High-Level Design (HLD) and Low-Level Design (LLD), aligned with NATO wargaming principles, enabling reproducibility and structured implementation in similar contexts.

\item The incorporation of a mathematical modeling and simulation core (based on the SERDUX-MARCIM model) into the adjudication process, allowing the representation of cyberattack propagation dynamics and the evaluation of defensive decisions in terms of network states and service availability.

\item A competency-oriented validation approach that goes beyond traditional subjective assessments, including a structured instrument to evaluate CSA-related learning outcomes and a comparative analysis against a control group to quantify the educational impact of the wargame.
\end{itemize}

The novelty of MARCIM-WG lies in the explicit integration of three dimensions that are rarely combined in existing cyber wargaming approaches: (i) the operationalization of Cyber Situational Awareness (CSA) as a central learning objective, rather than as an implicit outcome; (ii) the use of an analytically-assisted adjudication mechanism grounded in a mathematical simulation model, which ensures consistency and traceability between player decisions and system-level effects; and (iii) the articulation of a hybrid environment that combines physical interaction (tabletop elements) with computational modeling to support decision-making under realistic constraints. Unlike many existing cyber wargames and simulation-based training environments that rely primarily on expert judgement or predefined scripts, MARCIM-WG provides a dynamic, model-driven representation of cyberattack propagation and response, enabling participants to explore cause–effect relationships in a structured and measurable manner.
\section{State of the art}\label{sota}

Cyber wargaming is commonly positioned as a bridge between strategic intent and operational decision-making in cyber contexts, enabling structured exploration of adversarial dynamics, constraints, and trade-offs. In this line, MITRE frames cyber wargaming as a mechanism to embed realistic organizational context into decision processes so that leaders can reason about impacts, priorities, coordination, and risk acceptance rather than focusing solely on technical artifacts \cite{bodeau2018cyber}. Doctrine-oriented contributions likewise emphasize that cyber wargames should translate strategic guidance into actionable courses of action, supported by purposeful adjudication, well-designed injects, and explicit learning objectives to produce defensible and reusable insights \cite{curry2018cyber}. From a methodological standpoint, foundational wargaming perspectives stress disciplined game design and execution to ensure internal coherence between decisions, friction, and consequences, which is especially relevant in cyber environments where uncertainty and adaptation dominate \cite{weiner1959the}.

In parallel, simulation-based training environments and serious games have been widely adopted in cybersecurity and cyberdefense education. These approaches include technical exercises such as capture-the-flag scenarios, red-team/blue-team simulations, and tabletop exercises (TTX), which enable participants to rehearse operational responses under controlled conditions. While these environments are effective for developing technical skills and coordination, they often focus on isolated aspects of cyber incidents and provide limited representation of system-level dynamics and strategic decision-making processes. Additionally, many of these approaches rely on predefined scenarios or expert-driven adjudication, which may reduce traceability between decisions and outcomes and limit reproducibility across different training sessions.

Despite these advantages, reported implementations repeatedly highlight limitations that constrain transferability and educational value: uneven methodological rigor, weak or inconsistent validation, and difficulties in aligning mechanics with measurable competence outcomes \cite{onduto2021implementing}. Educational wargaming research and broader game-based learning literature reinforce that effectiveness depends on clear mapping between game activities and intended learning constructs, as well as the availability of instruments that can assess changes in participant competencies beyond subjective impressions \cite{mayer2016the}. These limitations motivate domain-grounded wargame designs that explicitly operationalize competence development and incorporate traceable adjudication and assessment mechanisms, particularly for specialized contexts where strategic cyber decision-making must be trained and evaluated against explicit targets.

Within this context, Cyber Situational Awareness (CSA) has been identified as a critical capability for effective decision-making in cyber environments. Based on the three-level model of perception, comprehension, and projection, CSA enables decision-makers to interpret technical signals, understand their implications in context, and anticipate possible future states of a cyber incident. Although CSA has been extensively studied in areas such as cyber monitoring, threat intelligence, and operational security, its explicit integration into cyber wargaming environments remains limited. In many existing wargames, situational awareness is treated as an implicit outcome of participation rather than as a structured objective supported by specific mechanics, decision cycles, and evaluation instruments.

Furthermore, in domain-specific contexts such as maritime cybersecurity, research has primarily focused on risk assessment, protection of critical infrastructures, and regulatory frameworks, with limited attention to training mechanisms that allow strategic-level actors to rehearse decision-making under evolving cyber crisis conditions. The complexity of maritime ecosystems, characterized by interconnected services, multiple stakeholders, and strong dependencies between cyber and physical systems, reinforces the need for approaches that integrate system dynamics with strategic decision processes.

Compared to existing cyber wargames and simulation-based training environments, several gaps can be identified. First, there is often a lack of formal integration between player decision-making and underlying system behavior, resulting in weak traceability between actions and outcomes. Second, the development of Cyber Situational Awareness is rarely addressed as a measurable and explicit objective. Third, validation approaches are frequently limited to qualitative observations, without structured instruments to assess competency development.

In this context, MARCIM-WG addresses these limitations by proposing a domain-grounded maritime cyberdefense wargame that integrates a hybrid tabletop environment with an analytically-assisted adjudication mechanism supported by a computational simulation model. Unlike traditional approaches, the proposal explicitly operationalizes CSA as a core learning objective, structures decision–consequence relationships through model-driven adjudication, and incorporates a competency-based validation strategy. This combination enables a more rigorous, traceable, and educationally grounded approach to cyber wargaming at the strategic level.
\section{Background}\label{background}

The \emph{NATO Wargaming Handbook} frames wargaming as a purpose-driven activity in which the game design, execution, and analysis are derived from a clearly stated intent (e.g., learning or analysis). From this perspective, a defensible wargame starts by articulating the problem to be explored, the objectives to be achieved, and the roles and responsibilities of the wargaming team and participants, ensuring that the overall setup (scenario, players, rules, and data capture) is coherent with the intended outcomes \cite{alliedcommandtransformation2023nato}. The methodology also emphasizes an end-to-end cycle that moves from design and development to execution and exploitation (after-action analysis and reporting), enabling refinement and institutional learning across iterations \cite{alliedcommandtransformation2023nato}.

A central contribution of the NATO approach is the operationalization of wargame design through four essential elements---\emph{decisions, friction, consequences, and narrative}. Together, these elements ensure that player choices are meaningful, constrained by realistic uncertainty and limitations, and connected to observable outcomes that can be discussed and exploited \cite{alliedcommandtransformation2023nato}. In addition, the handbook distinguishes adjudication modalities and recommends \emph{analytically-assisted} adjudication when consistency and traceability between decisions and effects are required, particularly in complex domains where outcomes must be interpreted against explicit objectives \cite{alliedcommandtransformation2023nato}. These principles provide the methodological basis adopted in this work to structure MARCIM-WG from high-level intent to low-level implementation.

The SERDUX-MARCIM model provides the mathematical and computational foundation for representing cyberattack dynamics in maritime environments. This model combines compartmental approaches inspired by epidemiological modeling with a cyber risk assessment structure to simulate the propagation of cyberattacks across interconnected systems. In particular, it extends classical models by incorporating time-dependent transition rates and variables associated with attacker capabilities, target conditions, and network characteristics, enabling a system-level representation of cyber incidents. The model captures the evolution of nodes through different states (e.g., susceptible, exposed, affected, and recovered), allowing the analysis of how attacks spread, how defensive actions influence system behavior, and how service availability is impacted over time \cite{cabuyapadilla2025serdux}. In the context of MARCIM-WG, this model is not used in its full analytical complexity; instead, it is operationally adapted to support the adjudication process, translating player decisions into parameter variations and producing quantitative outputs that reflect the evolving state of the simulated maritime cyber crisis.

\section{Proposal}
\label{sec:proposal}

\subsection{Overview of MARCIM-WG}
\begin{figure}[]
\centering
\includegraphics[width=\columnwidth]{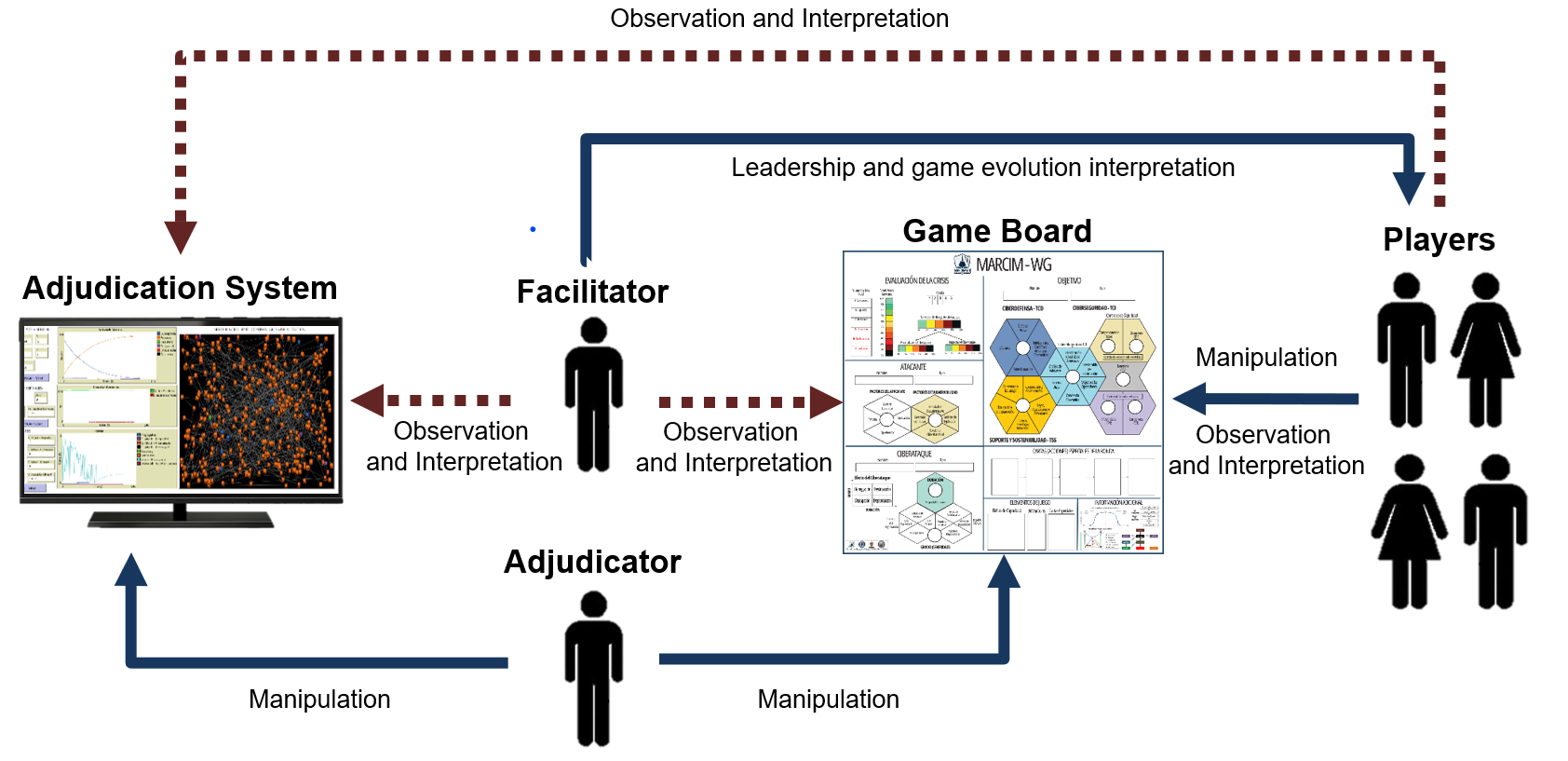}
\caption{MARCIM-WG interaction loop model between players, Game Board, and the analytically-assisted adjudication system.}
\label{fig:dynamics}
\end{figure}

MARCIM-WG \cite{cabuyapadilla2025marcimwg} is a learning-oriented maritime cyberdefense wargame that enables strategic-level participants to rehearse decision-making during a cyber crisis affecting a maritime actor. The proposal combines a physical tabletop (board, tokens, indicators, and special cards) with an analytically-assisted adjudication system implemented in software, so that player choices are translated into model inputs and returned as quantitative outputs that are interpreted collaboratively during facilitation. As illustrated in Fig.~\ref{fig:dynamics}, MARCIM-WG operationalizes a closed decision--adjudication--feedback loop: players decide and allocate resources on the tabletop, the facilitator/adjudicator encodes those decisions into the software, the simulation produces system-level effects, and the results are fed back to update the game state and structure the after-action discussion.

The wargame is conceived as a continuation and derived product of the MARCIM project \cite{cabuyapadilla2024marcimwg}, building on its maritime cyberdefense modeling-and-simulation foundation. It operationalizes CSA \cite{endsley1995toward,franke2014cyber} development through structured rounds, escalating threat dynamics, and explicit linkages between actions and consequences.

\subsection{High-Level Design Specifications (HLD)}
\subsubsection{Theme and problem framing}
MARCIM-WG is framed within maritime cyberdefense as a strategic and operational capacity to protect maritime power against cyber incidents. The design responds to a capability gap identified at the strategic level: limited appropriation of incident-response procedures and protocols in maritime cyber crisis contexts, compounded by increasing threat sophistication and the scarcity of specialized training tools for decision-makers in this domain \cite{cabuyapadilla2024marcimwg}.

\subsubsection{Wargame type and objectives}
Following the NATO classification, MARCIM-WG is conceived as a learning wargame focused on competence development rather than on validating a specific operational plan \cite{alliedcommandtransformation2023nato}. Its general objective is to provide a structured environment where participants analyze the propagation of a cyberattack over a critical maritime network and assess strategic response options. The specific objectives align with the CSA model by targeting: (i) \emph{perception} (identifying relevant cues, anomalies, and milestones), (ii) \emph{comprehension} (interpreting implications and vulnerabilities), and (iii) \emph{projection} (anticipating evolution and planning mitigations) \cite{endsley1995toward,franke2014cyber}.

\subsubsection{Core design elements}
The HLD explicitly operationalizes the four essential elements recommended by NATO \cite{alliedcommandtransformation2023nato}:
\begin{itemize}
  \item \textbf{Decisions:} players select response actions and allocate limited resources, exploring alternative courses of action under pressure.
  \item \textbf{Friction:} the design introduces constraints, uncertainty, and competing priorities through the game mechanics and role interactions.
  \item \textbf{Consequences:} decisions generate measurable effects on the maritime network and service levels through the adjudication system, enabling structured reflection.
  \item \textbf{Narrative:} a coherent crisis storyline frames the decision context and supports progressive escalation across rounds.
\end{itemize}

\subsubsection{Adjudication approach}
MARCIM-WG adopts an analytically-assisted adjudication method. Player decisions are not resolved only by expert judgement; instead, they are encoded as configurable inputs to a computational model that simulates their effects and returns quantitative outputs (e.g., evolution of node states and service availability) to be interpreted in the game context. Concretely, the adjudication software is implemented as an operational adaptation of the SERDUX-MARCIM simulation model \cite{cabuyapadilla2025serdux}, providing traceability between decisions, model execution, and observed effects on services and nodes. This approach increases consistency and supports results-oriented after-action analysis focused on CSA.

\subsubsection{Constraints and assumptions}
At high level, the design acknowledges constraints that condition fidelity and scalability, including computational requirements for dynamic-rate simulation, limits in the number of nodes/links that can be represented, fixed network structure during execution, and abstraction of maritime ecosystem elements to maintain playability and technical feasibility. The design also recognizes limitations associated with data availability for parametrization and the need to interpret results as decision-support learning outputs rather than predictive operational forecasts.

\subsubsection{Wargaming team and deliverables}
The HLD defines a multidisciplinary team organized around four methodological phases---design, development, execution, and analysis/reporting---consistent with NATO guidance \cite{alliedcommandtransformation2023nato}. Roles include sponsor, game director, designer, facilitator, adjudicator, and analyst functions. The expected deliverable includes a wargame report capturing key findings and lessons to inform future iterations and institutional learning.

\subsection{Low-Level Design Specifications (LLD)}
The LLD operationalizes the HLD into detailed rules, mechanics, physical components, and computational integration. The central product is the "Wargame Book", which structures the game content into: rules (including an ethical/operational confidentiality rule), general characteristics, objectives, roles, actor typology in the narrative, game elements (board, capability tokens, level indicators, BitMarCoins, special cards, dice, and adjudication system), scenario intent and phases, execution phases (preparation; execution across rounds; closure and evaluation), facilitator/adjudicator-only information, and a glossary.

\begin{figure}[]
\centering
\includegraphics[width=\columnwidth]{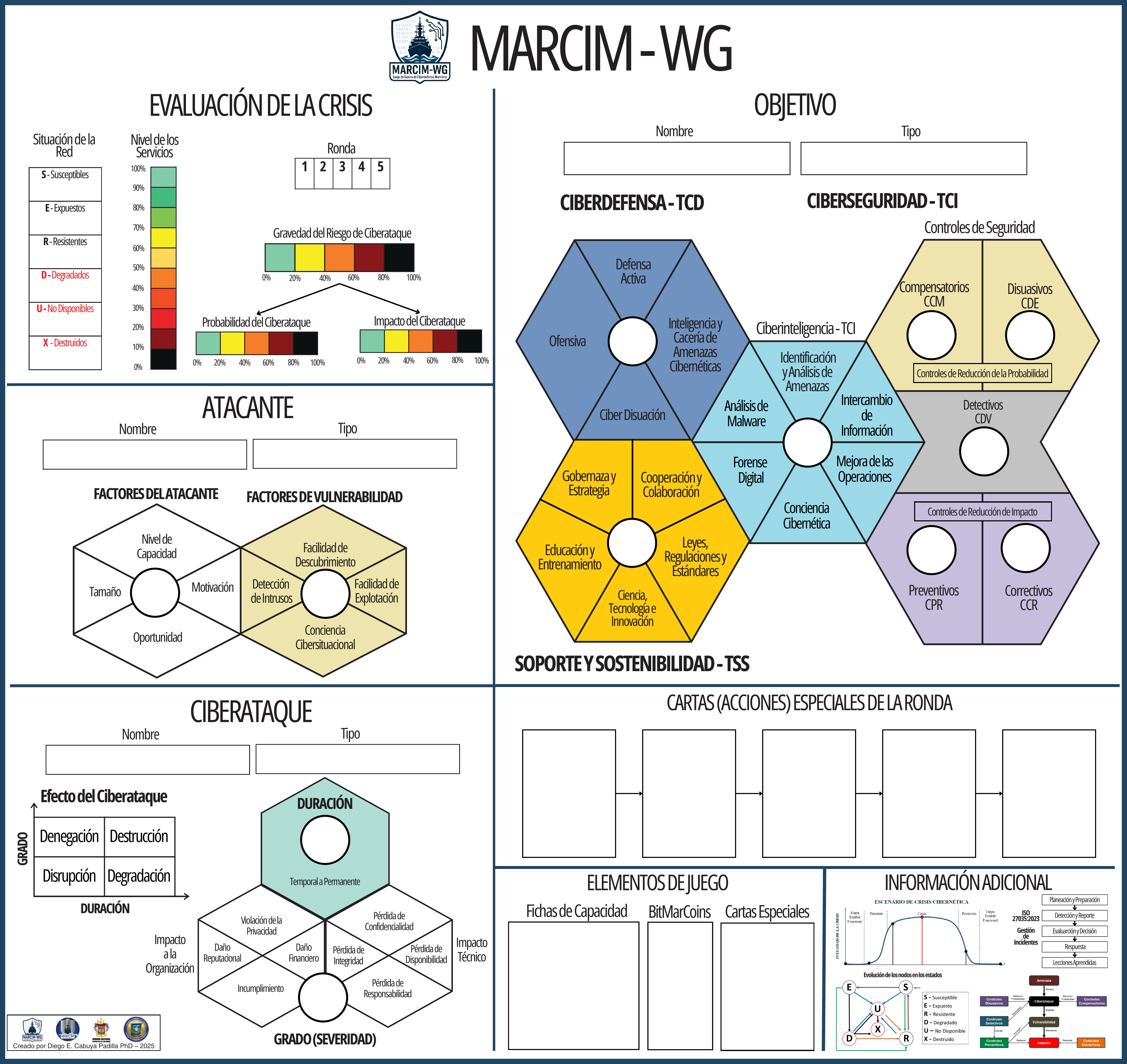}
\caption{Physical MARCIM-WG gameboard used by participants during gameplay to support decision-making, resource allocation, and state updates across adjudication cycles \cite{cabuyapadilla2025marcimwg}.}
\label{fig:board}
\end{figure}

The physical representation of the system under study is materialized through the MARCIM-WG physical gameboard (Fig.~\ref{fig:board}), which provides a shared visual language to manage state evolution, resources, and decisions during play. The board is updated after each adjudication cycle to reflect the new conditions of the crisis and to sustain discussion and sensemaking at the strategic level.

\subsubsection{Adjudication system}

\begin{figure}[]
\centering
\includegraphics[width=\columnwidth]{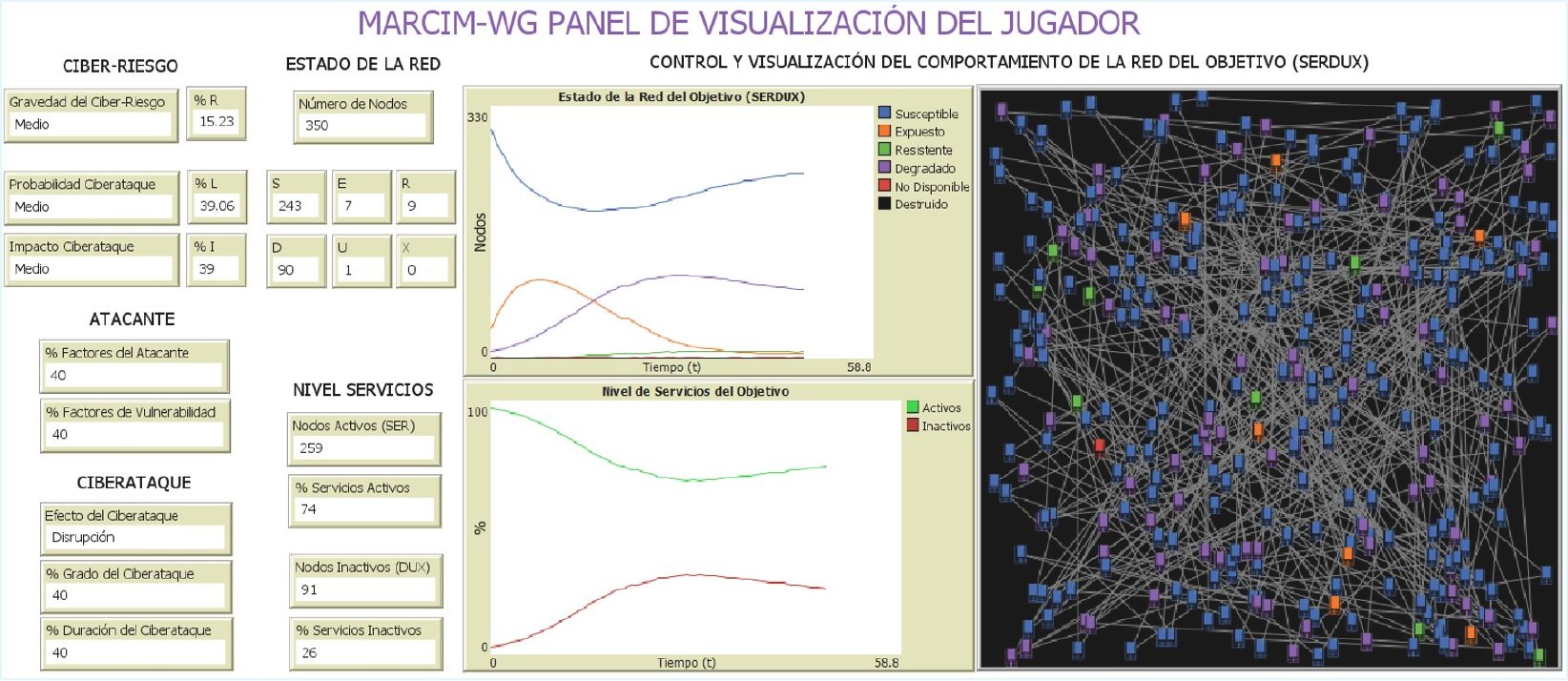}
\caption{MARCIM-WG adjudication software \cite{cabuyapadilla2025marcimwg} (SERDUX-MARCIM operational adaptation) and example of simulation outputs used for tabletop updates.}
\label{fig:adjudication}
\end{figure}

The adjudication system is implemented via the MARCIM-WG software (Fig.~\ref{fig:adjudication}) as an operational adaptation of the SERDUX-MARCIM model \cite{cabuyapadilla2025serdux}. The software is developed in NetLogo \cite{wilensky1999netlogo, wilensky2015an} with integrated Python \cite{pythonsoftwarefoundationn.d.python} modules, enabling the facilitator/adjudicator to input player decisions, execute simulation runs, and update the tabletop based on model outputs. The LLD also formalizes a competence-based learning structure: three CSA-oriented competences (perception, comprehension, projection) with observable dimensions, associated learning outcomes, and a pre/post instrument to compare initial and achieved levels \cite{endsley1995toward,paul2013a}.

To support the adjudication process, the underlying SERDUX-MARCIM model is operationally adapted from its original formulation into a decision-support structure that preserves its system-level dynamics while reducing analytical complexity for gameplay purposes. The model is based on a compartmental representation of cyberattack propagation, where network nodes evolve through different states over time according to transition rates influenced by attacker behavior, target conditions, and defensive actions. In MARCIM-WG, this model is integrated through a hierarchical organization of variables derived from the cyber risk approach defined in SERDUX-MARCIM \cite{cabuyapadilla2025serdux}. Specifically, the implementation considers variables up to level 3 of the hierarchy (Table~\ref{tab:variables_serdux}), where level 0 defines the overall cyber attack severity, level 1 separates likelihood and impact dimensions, and level 2 and level 3 provide progressively detailed factors associated with attacker behavior, target characteristics, defensive controls, and impact assessment. 

Within this structure (Table~\ref{tab:variables_serdux}), player decisions are mapped into modifications of selected level 3 variables, which act as operational inputs to the simulation. It is important to note that, although the original model defines a complete set of variables at this level, the MARCIM-WG implementation excludes specific elements that are not directly relevant to the gameplay abstraction, such as target network traffic, while preserving the remaining factors to ensure representativeness of cyber risk dynamics. During each adjudication cycle, the facilitator encodes player decisions into these variables, executes the simulation, and obtains quantitative outputs such as node-state distributions and service availability levels. These outputs are then interpreted and fed back into the tabletop, ensuring a consistent decision–model–feedback loop that enables participants to observe the systemic consequences of their actions in a dynamic cyber crisis scenario.

\begin{table*}[ht]
\centering
\caption{Hierarchical organization of variables adapted from SERDUX-MARCIM}
\label{tab:variables_serdux}
\begin{tabular}{|p{2cm}|p{4cm}|p{4cm}|p{6cm}|}
\hline
\textbf{Level 0} & \textbf{Level 1} & \textbf{Level 2} & \textbf{Level 3} \\ \hline

Cyber Risk 
& Cyber Attack Likelihood 
& Attacker Likelihood 
& Attacker Factors (ATF), Vulnerability Factors (VUF) \\ \cline{3-4}

& 
& Target Likelihood 
& Likelihood Reduction Controls ($\iota$), Target Cyber Intelligence Capability (TCI), Target Cyber Defense Capability (TCD), Target Support and Sustainability Capability (TSS) \\ \cline{2-4}

& Cyber Attack Impact 
& Impact Reduction Controls 
& Preventive Controls ($\theta$), Corrective Controls ($\eta$), Detective Controls ($\zeta$) \\ \cline{3-4}

& 
& Cyber Attack Degree 
& Technical Impact ($\tau$), Business Impact ($\rho$) \\ \cline{3-4}

& 
& Cyber Attack Duration 
& --- \\ \hline

\end{tabular}
\end{table*}

\subsubsection{Wargame Operational Dynamics}
To provide a clearer understanding of the operational dynamics of MARCIM-WG, the wargame execution is structured into three main phases: preparation, execution, and closure. During the preparation phase, the facilitator defines the scenario configuration, initializes the system state, and distributes initial resources (e.g., capability tokens) among participants to ensure a common baseline for decision-making. This phase also includes the briefing of participants regarding objectives, roles, and rules of interaction.

The execution phase is organized into iterative rounds, each following a consistent decision–adjudication–feedback cycle. In each round, participants assume the role of strategic decision-makers responsible for allocating resources, selecting response actions, and activating available game elements such as capability tokens, BitMarCoins, and special cards. These decisions are made under conditions of uncertainty, resource constraints, and evolving threat dynamics, reflecting the complexity of real-world cyber crisis management.

Once decisions are defined, the facilitator and adjudicator encode them into the simulation environment, where they are translated into parameter modifications of the underlying model. The simulation is then executed, producing quantitative outputs that represent the evolution of the cyberattack, the state of network nodes, and the level of service availability. These results are subsequently interpreted and used to update the gameboard, ensuring that players receive immediate and consistent feedback on the consequences of their actions.

The closure phase focuses on structured reflection and evaluation. During this phase, the outcomes of the simulation are analyzed, and participants engage in a guided discussion to assess the effectiveness of their decisions, identify strengths and weaknesses, and relate the observed dynamics to real-world cyberdefense procedures and protocols. Additionally, the competence-based assessment instrument is applied to evaluate the development of Cyber Situational Awareness (CSA) across its three dimensions: perception, comprehension, and projection.

From a role perspective, MARCIM-WG involves four main functions: (i) players, who act as strategic decision-makers; (ii) the facilitator, who guides the process and ensures adherence to the rules; (iii) the adjudicator, who manages the interaction with the simulation model; and (iv) the system itself, which represents the operational environment and enforces the consequences of decisions. This role distribution ensures a clear separation between decision-making, execution, and evaluation processes.

Overall, the structured sequence of phases and roles enables a coherent and reproducible gameplay experience, where decisions are systematically linked to observable outcomes. This design reinforces the learning objectives of the wargame by making explicit the relationship between strategic actions and their systemic effects, thereby supporting the development of Cyber Situational Awareness in complex environments.

\subsection{Application to a Maritime Cyberdefense Crisis Scenario}

The proposal is instantiated through a fully fictional maritime cyberdefense crisis scenario that provides narrative and technical coherence for gameplay. The scenario is dynamic: its trajectory depends on participant decisions under uncertainty and strategic pressure, while the adjudication system computes consequences in terms of cyberattack progression, impact on the target network, and institutional response capacity \cite{cabuyapadilla2024marcimwg,cabuyapadilla2025serdux}.

Participants influence the scenario through round-based control elements: (i) capability tokens allocated across the target's defensive capacities, (ii) BitMarCoins representing a constrained resource associated with crisis-specific decisions, and (iii) special cards that introduce discrete actions and structured frictions. In the first round, the facilitator distributes capability tokens uniformly across the target capacities to support familiarization; in subsequent rounds, allocation choices and the activation of additional elements depend on the players' strategic decisions. To reflect escalation, attacker and cyberattack parameters increase progressively across rounds, enabling the model to represent a transition from an initial threat into destructive effects executed by a sophisticated adversary. The scenario configuration includes initial network conditions and key parameters governing state evolution (e.g., propagation and resilience-loss dynamics), so that repeated decision--adjudication cycles expose players to measurable cause--effect relationships that support CSA development and structured after-action reflection.
\section{Validation and Results}\label{method}

To verify the formative impact and functional coherence of MARCIM-WG, two complementary validations were conducted. First, an \emph{conceptual validation} assessed whether the wargame and its adjudication system respond coherently to different decision patterns under three scenario configurations (pessimistic, neutral/most likely, and optimistic). Second, a \emph{CSA competency and learning-outcome validation} evaluated the extent to which MARCIM-WG contributes to strategic Cyber Situational Awareness (CSA) development through a structured assessment instrument and a comparative strategy with an equivalent control group.

\subsection{Conceptual validation (scenario-based execution)}
\label{subsec:operational_validation}
\begin{figure}[!h]
\centering
\includegraphics[width=\columnwidth]{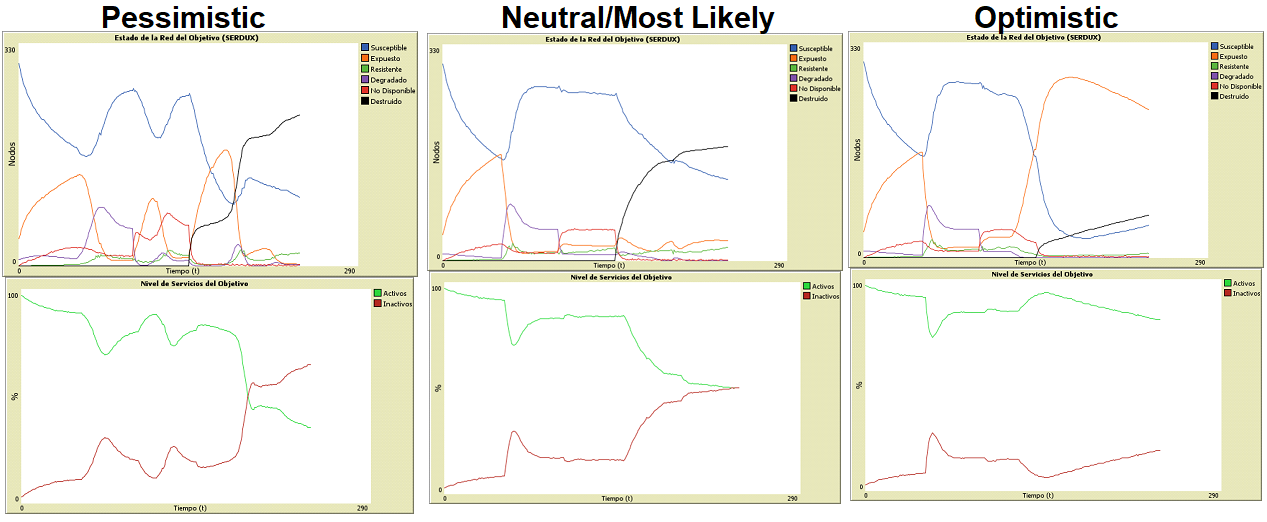}
\caption{Operational validation of MARCIM-WG through three scenarios (pessimistic, neutral/most likely, optimistic)\cite{cabuyapadilla2025marcimwg}.}
\label{fig:validation_operational}
\end{figure}

Conceptual validation was conducted by executing MARCIM-WG under three scenario configurations---pessimistic, neutral (most likely), and optimistic---to assess whether the adjudication logic and the resulting system behavior are sensitive to different strategic decision patterns. Fig.~\ref{fig:validation_operational} summarizes the scenario-based validation approach used during the operational trials. The outcomes were generated through the analytically-assisted adjudication process (operational adaptation of the SERDUX-MARCIM model) and interpreted in terms of service availability and network degradation.

The pessimistic scenario reflected limited planning and delayed or erratic use of special cards, resulting in rapid system degradation, a drop of service level below 35\%, and 224 destroyed nodes, with no effective recovery. The neutral (most likely) scenario represented a tactical response without strategic articulation; despite using special cards, actions were applied in a poorly coordinated manner and the system maintained around 50\% of active services, but with 175 destroyed nodes and a degraded network. In contrast, the optimistic scenario combined structured planning, timely activation of special cards, and coherent investment of capability tokens, preserving more than 80\% of active services and limiting destruction to 65 nodes. Collectively, these results support that MARCIM-WG is decision-sensitive and that the integration of narrative, physical resources, and computational adjudication yields coherent dynamics for training and discussion.

\subsection{CSA competency and learning-outcome validation}
\label{subsec:csa_validation}

\subsubsection{Purpose and validation design}
\label{subsec:validation_design}
To empirically verify the contribution of MARCIM-WG to CSA development, a comparative validation strategy was implemented using an equivalent control group. The purpose was to contrast the appropriation of key CSA concepts between participants who did not play the wargame and those who completed a controlled MARCIM-WG session. The validation followed a basic quasi-experimental design with post-test measurement between groups, treating the wargame experience as the independent variable and the measured competency attainment as the dependent variable.

\subsubsection{Participants profile}
\label{subsec:participants_profile}
Both the control group and the intervention group were composed of active-duty Colombian Armed Forces officers, selected to ensure technical-operational equivalence and a comparable strategic background. The selection criteria included: (i) institutional affiliation as active officers (Navy, Army, or Air Force); (ii) at least 20 years of professional experience in operational, strategic, or institutional support roles; and (iii) prior formal education (any level) related to cybersecurity, cyberdefense, or military strategy. Additional criteria ensured familiarity with cyber risk and incident management, as well as functional understanding of maritime power (doctrine, planning, or professional experience). This profile sought to guarantee that participants had the conceptual and strategic baseline required to interact meaningfully with the wargame.

\subsubsection{Competency and learning-outcome assessment instrument}
\label{subsec:assessment_instrument}
The assessment instrument was designed as a formative tool with a comparative function to measure the development (or improvement) of the three CSA-oriented competencies targeted by MARCIM-WG, aligned with the three CSA phases: perception, comprehension, and projection \cite{endsley1995toward,paul2013a}. It consists of 25 questions mapped to the three competencies and their corresponding learning outcomes (LOs), enabling aggregation at two levels: (i) overall performance and (ii) performance by competency and LO. While the instrument is suitable for pre/post applications, the present validation implemented it as a post-test comparison between equivalent groups (control vs.\ intervention) to isolate the effect of the wargame within the constraints of the validation session.

\subsubsection{Results and discussion}
\label{subsec:validation_results}
\begin{figure}[!h]
\centering
\includegraphics[width=\columnwidth]{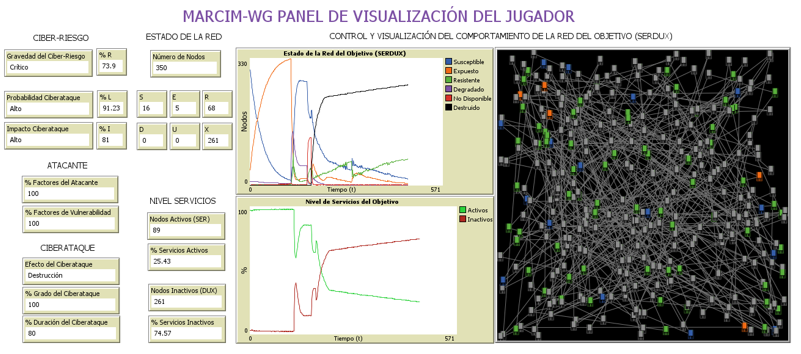}
\caption{Competency and learning-outcome validation of MARCIM-WG (CSA-focused assessment)\cite{cabuyapadilla2025marcimwg}.}
\label{fig:validation_csa}
\end{figure}

The competency validation session was conducted on July 10, 2025, at the Escuela Superior de Guerra ``General Rafael Reyes Prieto'' (Bogotá D.C., Colombia). First, the instrument was applied to a control group of eight officers who had not participated in MARCIM-WG. Subsequently, four officers with the same profile completed a controlled MARCIM-WG session and answered the same instrument immediately after finishing the game. Fig.~\ref{fig:validation_csa} depicts the competency-validation session context associated with this comparative assessment.

Table~\ref{tab:validation_results} summarizes the comparative results. At the global level, the intervention group achieved 91.2\% correct answers versus 57.2\% in the control group, yielding a reported improvement of +34.0 percentage points. By competency, the largest gain was reported for Competency~2 (integrated comprehension of cyber-crisis scenarios), followed by Competency~1 (strategic perception of the environment), and Competency~3 (strategic foresight - future scenarios).

\begin{table}[!h]
\centering
\caption{Comparative results of competency and learning-outcome validation (control vs.\ intervention).}
\label{tab:validation_results}
\scriptsize
\renewcommand{\arraystretch}{1}
\setlength{\tabcolsep}{0pt}
\begin{tabular}{@{}
>{\raggedright\arraybackslash}p{0.57\columnwidth}
@{\hspace{6pt}}
>{\raggedleft\arraybackslash}p{0.1\columnwidth}
@{\hspace{6pt}}
>{\raggedleft\arraybackslash}p{0.1\columnwidth}
@{\hspace{6pt}}
>{\raggedleft\arraybackslash}p{0.1\columnwidth}
@{}}
\hline
\textit{Competency / LO} & \textit{Ctrl} & \textit{Intv} & \textit{Diff} \\
\hline
Overall result & 57.2 & 91.2 & +34.0 \\
\hline
Competency 1 -- Strategic perception & 66.6 & 88.8 & +29.2 \\
\hspace{0.6em}LO 1.1 Identify critical assets & 70.0 & 100.0 & +30.0 \\
\hspace{0.6em}LO 1.2 Recognize active threats & 54.3 & 100.0 & +45.7 \\
\hspace{0.6em}LO 1.3 Assess cyber risk level & 54.6 & 66.6 & +12.0 \\
\hline
Competency 2 -- Integrated comprehension & 61.9 & 97.5 & +35.54 \\
\hspace{0.6em}LO 2.1 Analyze attack evolution & 71.8 & 100.0 & +28.2 \\
\hspace{0.6em}LO 2.2 Evaluate cyber capabilities & 44.0 & 100.0 & +56.0 \\
\hspace{0.6em}LO 2.3 Recognize crisis phases & 71.5 & 100.0 & +28.5 \\
\hspace{0.6em}LO 2.4 Apply ISO/IEC 27035 principles & 60.0 & 100.0 & +40.0 \\
\hspace{0.6em}LO 2.5 Analyze strategic implications & 62.5 & 87.5 & +25.0 \\
\hline
Competency 3 -- Strategic foresight & 78.3 & 89.5 & +11.2 \\
\hspace{0.6em}LO 3.1 Formulate strategic hypotheses & 87.5 & 100.0 & +12.5 \\
\hspace{0.6em}LO 3.2 Evaluate effects of decisions & 59.3 & 68.7 & +9.3 \\
\hspace{0.6em}LO 3.3 Reflect on operational/doctrinal implications & 88.0 & 100.0 & +12.0 \\
\hline
\multicolumn{4}{@{}p{\columnwidth}@{}}{\footnotesize Source: own elaboration based on the validation session results \cite{cabuyapadilla2025marcimwg}.} \\
\end{tabular}
\end{table}

At the LO level, all measured outcomes improved, with reported gains ranging from +9.3 to +56.0 points. The largest single improvement corresponded to LO~2.2 (evaluation of cyber capabilities), rising from 44.0\% to 100.0\% (+56.0). Other notable gains included LO~1.2 (recognition of active threats), from 54.3\% to 100.0\% (+45.7), and LO~2.4 (application of ISO/IEC~27035 principles), from 60.0\% to 100.0\% (+40.0). The lowest LO-level gain was observed in LO~3.2 (evaluation of decision effects), from 59.3\% to 68.7\% (+9.3), suggesting that projection-oriented reasoning may require additional repetitions, more time for reflection, or complementary instructional scaffolding.

Overall, the CSA-focused results show that MARCIM-WG strengthens perception, comprehension, and projection, with strong effects on comprehension-related outcomes. However, the evidence should be interpreted with caution due to the limited sample size and post-test-only, between-groups design; future replications with larger cohorts would allow stronger inference and clearer characterization of retention.
\section{Conclusions}\label{conclusions}

This paper presented MARCIM-WG, a learning-oriented maritime cyberdefense wargame designed to strengthen strategic Cyber Situational Awareness (CSA) through a hybrid approach that combines a physical tabletop experience with analytically-assisted adjudication supported by computational simulation. The proposal operationalizes meaningful decision cycles under friction, links actions to measurable consequences, and leverages a coherent narrative to support structured reflection and institutional learning.

The validation results indicate that MARCIM-WG contributes to CSA development across the three targeted competencies (perception, comprehension, and projection). The most substantial improvements were observed in comprehension-related outcomes, particularly those associated with evaluating cyber capabilities and interpreting crisis dynamics, which suggests that the wargame effectively supports higher-order reasoning that connects technical signals with strategic decision-making. At the same time, the comparatively smaller gains in projection-oriented outcomes reinforce the need for repeated gameplay iterations and additional instructional scaffolding to consolidate anticipatory reasoning in complex cyber crises.

A key practical contribution of MARCIM-WG is that the adjudication system enables the use of a rigorous mathematical simulation model without transferring its full complexity to the players. By encapsulating the underlying dynamics within the adjudication software---and exposing participants only to decision levers and interpretable outputs---the approach proved viable for training and assessment purposes. This design choice also indicates broader applicability: the same adjudication logic, grounded on a compartmental and agent-based simulation core, can be adapted to non-maritime cybersecurity and cyberdefense scenarios by modifying the scenario configuration, network representation, and decision levers while preserving the decision--consequence structure of the wargame.

The main limitations of this study are the reduced sample size and the post-test-only comparison design. Future work should replicate the assessment with larger cohorts, incorporate repeated-measures designs to evaluate retention over time, and refine projection-oriented learning elements (e.g., additional injects, longer decision windows, and more structured after-action prompts) to improve anticipatory reasoning and transfer of learning.

Supporting documentation, scenario assets, datasets, and implementation are available at \url{https://github.com/diegocabuya/SERDUX-MARCIM}.

%%referencias
\bibliographystyle{IEEEtran}
\bibliography{bibliography.bib}{}

@article{alcaide2020critical,
  author  = {Alcaide, J. I. and Llave, R. G.},
  title   = {Critical infrastructures cybersecurity and the maritime sector},
  journal = {Transportation Research Procedia},
  year    = {2020},
  volume  = {45},
  pages   = {547--554},
  doi     = {10.1016/j.trpro.2020.03.058}
}

@misc{alliedcommandtransformation2023nato,
  author = {{Allied Command Transformation}},
  title  = {NATO wargaming handbook},
  year   = {2023},
  howpublished = {\url{https://paxsims.wordpress.com/wp-content/uploads/2023/09/nato-wargaming-handbook-202309.pdf}}
}

@techreport{bodeau2018cyber,
  author = {Bodeau, D. J. and Mccollum, C. D. and Fox, D. B.},
  title  = {Cyber wargaming: framework for enhancing cyber wargaming with realistic business context},
  year   = {2018},
  institution ={MITRE},
  note   = {\url{http://www.mitre.org/HSSEDI}}
}

@phdthesis{cabuyapadilla2024marcimwg,
  author = {Cabuya-Padilla, D. E.},
  title  = {Marco de Referencia para el Modelamiento y Simulaci{\'o}n de la Ciberdefensa Mar{\'\i}tima a Nivel Estrat{\'e}gico -- MARCIM},
  school = {Escuela Naval de Cadetes ``Almirante Padilla''},
  year   = {2024},
  note   = {[Tesis Doctoral]}
}

@article{cabuyapadilla2024ciberseguridad,
  author  = {Cabuya-Padilla, D. E. and Alvarado Carvajal, C. F. and Carrascal Ortiz, R. A. and Riola Rodr{\'\i}guez, J. M. and Fajardo-Toro, C. H. and Escandon Bernal, S. P.},
  title   = {Ciberseguridad y ciberdefensa mar{\'\i}tima: an{\'a}lisis bibliom{\'e}trico a{\~n}os 1990 -- 2021},
  journal = {RISTI - Revista Iberica de Sistemas e Tecnologias de Informacao},
  year    = {2022},
  volume  = {49},
  pages   = {197--210},
  howpublished = {\url{https://www.risti.xyz/issues/ristie49.pdf}}
}

@inproceedings{cabuyapadilla2025hybrid,
  author    = {Cabuya-Padilla, D. E. and D{\'\i}az-L{\'o}pez, D. and Castaneda-Marroquin, C.},
  title     = {Hybrid Tabletop Exercise (TTX) based on a Mathematical Simulation-based Model for the Maritime Sector},
  booktitle = {Actas de las X Jornadas Nacionales de Investigaci{\'o}n en Ciberseguridad: Zaragoza, 4 a 6 de junio de 2025},
  year      = {2025},
  pages     = {37--44},
  note      = {\url{https://2025.jnic.es/}}
}

@article{cabuyapadilla2025serdux,
  author  = {Cabuya-Padilla, D. E. and D{\'\i}az-L{\'o}pez, D. and Mart{\'\i}nez-P{\'a}ez, J. and Hern{\'a}ndez, L. and Castaneda-Marroquin, C.},
  title   = {SERDUX-MARCIM: Maritime Cyberattack simulation using Dynamic Modeling, Compartmental Models in Epidemiology and Agent-based Modeling},
  journal = {International Journal of Information Security},
  year    = {2025},
  volume  = {24},
  number  = {3},
  pages   = {122},
  doi     = {10.1007/s10207-025-00985-6}
}

@article{cabuyapadilla2024dyna,
  author  = {Cabuya-Padilla, D. E. and Castaneda-Marroquin, C. A.},
  title   = {Marco de referencia para el modelamiento y simulaci{\'o}n de la ciberdefensa mar{\'\i}tima - MARCIM: estado del arte y metodolog{\'\i}a},
  journal = {DYNA},
  year    = {2024},
  volume  = {91},
  number  = {231},
  pages   = {169--179},
  doi     = {10.15446/dyna.v91n231.109774}
}

@inproceedings{curry2018cyber,
  author    = {Curry, J. and Drage, N.},
  title     = {Developments in state level cyber wargaming},
  booktitle = {ACM International Conference Proceeding Series},
  year      = {2018},
  doi       = {10.1145/3264437.3264468}
}

@article{endsley1995toward,
  author  = {Endsley, M. R.},
  title   = {Toward a theory of situation awareness in dynamic systems},
  journal = {Human Factors},
  year    = {1995},
  volume  = {37},
  number  = {1},
  pages   = {32--64}
}

@article{franke2014cyber,
  author  = {Franke, U. and Brynielsson, J.},
  title   = {Cyber situational awareness--a systematic review of the literature},
  journal = {Computers \& Security},
  year    = {2014},
  volume  = {46},
  pages   = {18--31}
}

@inproceedings{jacq2019cyber,
  author    = {Jacq, O. and Brosset, D. and Kermarrec, Y. and Simonin, J.},
  title     = {Cyber attacks real time detection: towards a cyber situational awareness for naval systems},
  booktitle = {2019 International Conference on Cyber Situational Awareness, Data Analytics and Assessment (Cyber SA)},
  year      = {2019},
  note      = {June 1},
  doi       = {10.1109/CyberSA.2019.8899351}
}

@article{karim2022maritime,
  author  = {Karim, M. S.},
  title   = {Maritime cybersecurity and the IMO legal instruments: Sluggish response to an escalating threat?},
  journal = {Marine Policy},
  year    = {2022},
  volume  = {143},
  pages   = {105138},
  doi     = {10.1016/j.marpol.2022.105138}
}

@incollection{mayer2016the,
  author    = {Mayer, I. and Warmelink, H. and Zhou, Q.},
  title     = {The utility of games for society, business, and politics},
  booktitle = {The Wiley Handbook of Learning Technology},
  year      = {2016},
  pages     = {406--435},
  PUBLISHER ={Wyley},
  doi       = {10.1002/9781118736494.ch22}
}

@article{mrakovi2019maritime,
  author  = {Mrakovi{\'c}, I. and Vojinovi{\'c}, R.},
  title   = {Maritime cyber security analysis -- how to reduce threats?},
  journal = {Transactions on Maritime Science},
  year    = {2019},
  volume  = {8},
  number  = {1},
  pages   = {132--139},
  doi     = {10.7225/toms.v08.n01.013}
}

@mastersthesis{onduto2021implementing,
  author = {Onduto, B.},
  title  = {Gamification of cyber security awareness -- A systematic review of games},
  school = {University of Turku},
  year   = {2021},
  note   = {Master of Science in Technology Thesis. Computing, Faculty of Technology},
  howpublished = {\url{https://www.utupub.fi/bitstream/handle/10024/152929/Onduto_Barack_Thesis_Final.pdf}}
}

@incollection{paul2013a,
  author    = {Paul, C. L. and Whitley, K.},
  title     = {A Taxonomy of Cyber Awareness Questions for the User-Centered Design of Cyber Situation Awareness},
  booktitle = {LNCS},
  year      = {2013},
  volume    = {8030},
  pages     = {145--154},
  publisher = {Springer, Berlin, Heidelberg},
  doi       = {10.1007/978-3-642-39345-7_16}
}

@misc{pythonsoftwarefoundationn.d.python,
  author = {{Python Software Foundation}},
  title  = {Python},
  year   = {2023},
  howpublished = {\url{https://www.python.org/}}
}

@article{symes2024the,
  author  = {Symes, S. and Blanco-Davis, E. and Graham, T. and Wang, J. and Shaw, E.},
  title   = {Cyberattacks on the maritime sector: a literature review},
  journal = {Journal of Marine Science and Application},
  year    = {2024},
  pages   = {1--18}
}

@techreport{unctad2024review,
  author      = {{United Nations Conference on Trade and Development - UNCTAD}},
  title       = {Review of maritime transport 2024},
  institution = {{United Nations Conference on Trade and Development. UNCTAD}},
  year        = {2024},
  howpublished = {\url{https://unctad.org/system/files/official-document/rmt2024_en.pdf}}
}

@misc{weiner1959the,
  author = {Weiner, M. G.},
  title  = {War gaming methodology},
  year   = {1959},
  doi    = {10.1163/9789087903107_006},
  howpublished = {\url{https://doi.org/10.1163/9789087903107_006}}
}

@misc{wilensky1999netlogo,
  author = {Wilensky, U.},
  title  = {NetLogo},
  year   = {2016},
  howpublished = {\url{https://ccl.northwestern.edu/netlogo/}}
}

@book{wilensky2015an,
  author    = {Wilensky, U. and Rand, W.},
  title     = {An Introduction to Agent-Based Modeling: Modeling Natural, Social, and Engineered Complex Systems with NetLogo},
  year      = {2015},
  publisher = {MIT Press}
}

@mastersthesis{cabuyapadilla2025marcimwg,
  author  = {Cabuya-Padilla, D. E.},
  title   = {{MARCIM-WG: Juego de guerra de ciberdefensa mar{\'\i}tima para la apropiaci{\'o}n estrat{\'e}gica de respuestas ante crisis cibern{\'e}ticas}},
  school  = {Escuela Superior de Guerra ``General Rafael Reyes Prieto''},
  year    = {2025},
  address = {Bogot{\'a} D.C., Colombia},
  type    = {Monograf{\'\i}a de maestr{\'\i}a},
  note    = {Maestr{\'i}a en Ciberseguridad y Ciberdefensa.}
}

\end{document}